\documentstyle[emulateapj]{article}
\input{psfig}

\begin{document}

\submitted{ApJ Lett. submitted}

\title{Two-dimensional galaxy-galaxy lensing: a direct measure of the
flattening and alignment of light and mass in galaxies}

\author{Priyamvada Natarajan$^{1,2}$ \& Alexandre Refregier$^{1}$}
\affil{1 Institute of Astronomy, Madingley Road, Cambridge CB3 0HA, UK}
\affil{2 Department of Astronomy, Yale University, New Haven, CT, USA}

\begin{abstract}
We propose a new technique to directly measure the shapes of dark
matter halos of galaxies using weak gravitational lensing. Extending
the standard galaxy-galaxy lensing method, we show that the shape
parameters of the mass distribution of foreground galaxies can be
measured from the two-dimensional shear field derived from background
galaxies. This enables the comparison of the ellipticity of the mass
distribution with that of the light in galaxies, as well as an
estimate of the degree of alignment between the stellar component and
dark matter. We choose the specific case of an elliptical, isothermal
profile and estimate the feasibility and significance of the detection
of this signal. The prospects for applying this technique are
excellent with large on-going surveys like the Sloan Digital Sky
Survey. The expected signal is smaller, but comparable in significance
to that of the mass in standard galaxy-galaxy lensing analyses. Since
shapes of halos depend on the degree of dissipation and the transfer
of angular momentum during galaxy assembly, constraints obtained from
the analysis will provide an important input to models of galaxy
formation.
\end{abstract}
 

\section{Introduction}

Current observations require the existence of dark matter halos for
galaxies. However, fundamental parameters such as their total mass and
spatial extent are not well constrained. The mass distribution in
galaxies is primarily probed via dynamical tracers of the galactic
potential on various scales: stars in the inner regions, HI gas in
regions outside the optical radius, and orbital motions of bound
satellites in the outer-most regions (e.g. Zaritsky \& White 1994).
Probes of halo structure at radii devoid of any luminous tracers are
therefore needed, weak gravitational lensing offers precisely that.

Galaxy-galaxy lensing, the preferential tangential alignment of the
images of background galaxies around bright foreground ones, is
detected statistically by stacking galaxies. The first observational
attempt to look for galaxy-galaxy lensing was made by Tyson et
al. (1984). Recent studies have been very successful, and a signal at
the 99.5\% confidence level was first reported by Brainerd, Blandford
\& Smail (1996) using deep ground-based CCD data. Several subsequent
studies using Hubble Space Telescope images and ground-based data
(Griffiths et al. 1996; Dell' Antonio \& Tyson 1996; Hudson et
al. 1998; Casertano 1999; Ebbels et al. 2000; Fischer et al. 1999
[Sloan Digital Sky Survey (SDSS hereafter) commissioning data]) report
unambiguous detection of a galaxy--galaxy lensing signal.

In the method proposed in this letter, additional information that is
available but not exploited in current galaxy-galaxy lensing studies
is utilized, namely, the light distribution of galaxies selected to be
foreground lenses. The general results derived by Schneider \&
Bartelmann (1997) are used to relate the shear field to the mass
multipole moments. We show that the shapes and orientations of the
foreground galaxies (probes of the light) can be compared
statistically to that of the shear field (probe of the mass), thus
providing a direct method to compare the ellipticity of the light to
that of the mass as well any potential misalignments between them
reliably.  These parameters offer an important clue to the galaxy
formation process, since they provide a quantitative measure of the
importance of dissipation in the assembly of galaxies. Any variation
of the flattening of the total mass (predominantly the dark matter
component) with radius is a probe of the efficiency of angular
momentum transfer to the dark halo and might provide insights into the
structure and composition of galaxy halos. With regard to the relative
orientation of the light and mass in galaxies, the two components are
likely to be aligned on average and any misalignments might occur
transiently, for instance, after a violent merger.
     
The outline of the paper is as follows: in \S\ref{current}, the current
status of modeling in galaxy-galaxy lensing studies is reviewed. In
\S\ref{mass_shape} the formalism used to extract the shape of the mass
distribution is presented, with the application to an elliptical
isothermal mass model in \S\ref{model}.  The feasibility of detection
of the flattening of the mass is examined in \S\ref{flattening}, and a
measure of the alignment of mass and light is discussed in
\S\ref{alignment}. We conclude in \S\ref{discussion} with a discussion
of the importance of studying shapes of dark halos for understanding
key issues in the formation and structure of galaxies and the
relation between the luminous and dark component in galaxies.

\section{Current Status of galaxy-galaxy lensing}
\label{current}
The galaxy-galaxy lensing signal is measured ideally from a deep image
by selecting a population of brighter (assumed to be foreground)
galaxies as lenses and measuring the induced shape distortion in the
fainter (background) galaxies. The shear signal obtained from direct
averaging in radial bins around the bright foreground galaxies is then
stacked to obtain the radial profile of the shear $\gamma$ as a
function of distance from the lens. This provides reasonable
constraints on the circular velocity of a fiducial halo, found to be
in the range 210 km s$^{-1}$ -- 250 km s$^{-1}$ for a typical $L^*$
galaxy, but is found to be fairly insensitive to the radial extent of
the halos (consistent with halo sizes in excess of 100h$^{-1}$
kpc). In analyses to date, the errors are dominated by shot noise and
error arising from the unknown redshift distribution of
galaxies. However, with forthcoming surveys like the SDSS, which plan
to image over a hundred million galaxies in many bands and provide
reliable photometric redshifts, the prospects for galaxy-galaxy
lensing studies are extremely good (Fischer et al. 1999).

\section{Extracting shape parameters for the mass distribution}
\label{mass_shape}
The mass distribution of a lensing galaxy is described by the
convergence field $\kappa({\mathbf x})$, defined to be the surface
mass-density $\Sigma({\mathbf x})$ expressed in units of the critical
surface mass density $\Sigma_{\rm crit}$.  The critical surface mass
density depends on the precise configuration, i.e. on the angular
diameter distances from the lens to the source $D_{\rm ls}$, observer
to source $D_{\rm os}$ and observer to lens $D_{\rm ol}$. It is given
by, ${\Sigma_{\rm crit}}\,=\, {{c^2}\over{4 \pi G}}\,{{D_{\rm
os}}\over{D_{\rm ls}\,D_{\rm ol}}}$. Standard galaxy-galaxy lensing
provides a measure of the mass $M$ within an aperture, which is given
by
\begin{equation}
\label{eq:m}
M = \int\,d^{2} x~ w(x) \kappa({\mathbf x}),
\end{equation}
where $w(x)$ is a weight function normalized so that $\int d^{2}x~
w(x) = 1$. It is chosen to be continuous, differentiable and is
required to fall-off rapidly to zero outside the aperture scale
$\beta$. The shape parameters of the mass distribution are
characterized by the quadrupole moments of the convergence
$\kappa({\mathbf x})$ within the aperture (Schneider \& Bartelmann
1997), which are defined as
\begin{eqnarray}
Q_{ij}\, \equiv \,\int\,d^2 x\,x_i\,x_j\,w(x)\,\kappa({\mathbf x}),
\end{eqnarray}
This tensor can be decomposed into a trace-free piece $Q$ and a trace $T$
defined as:
\begin{eqnarray}
\label{eq:q_t}
Q\,=\,Q_{11}\,-\,Q_{22}\,+\,2\,i\,Q_{12},\,\,\,T\,=\,Q_{11}\,+\,Q_{22}.
\end{eqnarray}
The ellipticity of the mass $\epsilon_{\kappa}$ is then simply,
\begin{eqnarray}
\label{eq:epsilon_k}
\epsilon_{\kappa}\,=\,{\frac{Q}{T}}\,=\,\frac{(a_{\kappa}^2 -
b_{\kappa}^2)}{(a_{\kappa}^2 + b_{\kappa}^2)} e^{i \varphi_{\kappa}},
\end{eqnarray}
where $a_{\kappa}$ and $b_{\kappa}$ are, respectively, the major and
minor axes of the mass distribution, and $\varphi_{\kappa}$ is its position
angle relative to the positive $x$-axis.  

Schneider \& Bartelmann (1997) have shown that the multipole moments
of $\kappa({\mathbf x})$ computed from the observed shear
$\gamma({\mathbf x})$ field. In particular, the quadrupole moments
(Eq.~[\ref{eq:q_t}]) can be expressed as
\begin{eqnarray}
\label{eq:q_shear}
Q\,=\,\int\,d^2
x\,e^{2i\varphi}\,\left[ g_{t}(x)\,\gamma_t({\mathbf x})\,
+ i\,g_{\times}(x)\,\gamma_{\times}({\mathbf x}) \right],
\end{eqnarray}
where the rotated shear components $\gamma_{t}$ and $\gamma_{\times}$
correspond, respectively, to a tangential and curl-type shear pattern
about the center of mass of the lens (see Rhodes, Refregier \& Groth
2000 for an illustration). They are related to the unrotated
components by,
\begin{eqnarray}
\gamma_{t} & = & - \left[ \cos(2\varphi) \gamma_{1} + \sin(2 \varphi)
\gamma_{2} \right] \nonumber \\
\gamma_{\times} & = & - \left[ -\sin(2\varphi) \gamma_{1} + \cos(2 \varphi)
\gamma_{2} \right]
\end{eqnarray}
where $\varphi$ is the polar angle from the $x$-axis. The associated
aperture functions $g_{t}(x)$ and $g_{\times}(x)$ are given by
\begin{eqnarray}
g_{t}(x) = 2 \,V_{2}(x)\,-\,x^2\,w(x), ~~
g_{\times}(x) = -2 V_{2}(x),
\end{eqnarray}
where $V_{n}(x) = x^{-2} \int_{0}^{x} dx'~ x^{\prime n+1} w(x')$.
Similarly, the trace part $T$ and the mass $M$ can also be written as
\begin{eqnarray}
\label{eq:t_m_shear}
T\,=\,\int\,d^2 x\, g_{t}(x)  \,\gamma_t({\mathbf x}), ~~
M\,=\,\int\,d^{2} x\, h_{t}(x) \,\gamma_t({\mathbf x}),
\end{eqnarray}
where $h_{t}(x)=2 V_{0}(x) - w(x)$.

\section{Application to the elliptical isothermal model}
\label{model}
We consider an isothermal model with concentric elliptical
equipotentials (Natarajan \& Kneib 1996). The projected potential for
this model is $\psi=\alpha r$, where $\alpha$ is the Einstein radius
and $r$ is a generalized elliptical radius.  If the $x$-axis is
aligned with the major axis of the potential, the generalized radius
is given by $r^{2} = \frac{x_{1}^{2}}{1+\epsilon} +
\frac{x_{2}^{2}}{1-\epsilon}$, where $\epsilon$ is the ellipticity of
the equipotentials. The Einstein radius is related to the velocity
dispersion of the galaxy $\sigma_{v}$ by $\alpha = 4 \pi
(\frac{\sigma_{v}}{c})^2 ( \frac{D_{\rm ls}}{D_{\rm os}})$, and is of
the order of $1''$ for galaxies.

For a weakly elliptical model ($\epsilon \ll 1$), the potential has
the form,
\begin{eqnarray}
\psi ({\mathbf x})\,\simeq \,\alpha\,x\,[1-{\frac{\epsilon}{2}}\,
\cos\,2(\varphi-\varphi_{\rm 0})];
\end{eqnarray}
where $\varphi_{0}$ is the position angle of the potential, and
reduces to that of a singular isothermal sphere in the circular limit
($\epsilon=0$). Current observational limits on ellipticities of halos
have been compiled in a comprehensive recent review by Sackett (1999),
however, there are no constraints on the dispersion in the shape
parameters, therefore, for the purposes of this calculation we have
used the central value of $\epsilon = 0.3$.  Besides, higher order
terms will be approximately $(0.3)^2 = 0.09$ which are still only
corrections at the 10\% level, consistent with the assumption of small
$\epsilon$.

Restricting our analysis to the weak regime, to first order in
$\epsilon$, the associated convergence $\kappa = \nabla^{2} \psi /2$,
where $\psi=\alpha r$, is given by,
\begin{eqnarray}
\kappa\,({\mathbf x})\, \simeq \,{\frac{\alpha}{2\,x}}\,
[1\,+\,{\frac{3 \epsilon}{2}}\,\cos\,2 (\varphi-\varphi_0)],
\end{eqnarray}
and the complex shear $\gamma = \gamma_1 + i\gamma_2 =
[ \partial_{1}^{2} -\partial_{2}^{2} + 2 i
\partial_{1}\partial_{2}  ] \psi/2$ is
\begin{eqnarray}
\gamma\,\simeq\,-{\frac{\alpha}{2\,x}}\,
[1\,+\,{\frac{3\,\epsilon}{2}} \cos 2(\varphi-\varphi_0)]\, e^{2i\varphi},
\end{eqnarray}
The rotated shear components are thus
\begin{eqnarray}
\gamma_t\,\simeq\,{\frac{\alpha}{2\,x}}\, [1\,+\,{\frac{3\,\epsilon}{2}}
\cos 2(\varphi-\varphi_0)], ~~ \gamma_{\times}=0,
\end{eqnarray}
yielding a tangential shear modulated by an elliptical pattern. 

The ellipticity of the underlying mass distribution $\kappa ({\mathbf
x})$ needs to be related to that of the projected (2-d) potential
$\phi({\mathbf x})$. Using a normalized gaussian as the weight
function $w(x) = e^{-x^2/2\beta^2}/(2 \pi \beta^{2})$, we evaluate the
integral for the quadrupole (Eq.~[\ref{eq:q_t}]) and monopole
(Eq.~[\ref{eq:m}]) moments of $\kappa({\mathbf x})$,
\begin{eqnarray}
M=\sqrt{\frac{\pi}{8}} \frac{\alpha}{\beta},~~
|Q|=\frac{3}{8}\sqrt{\frac{\pi}{2}} \alpha \beta \epsilon,~~
T=\sqrt{\frac{\pi}{8}}  \alpha \beta
\end{eqnarray}
The ellipticity of the mass (Eq.~[\ref{eq:epsilon_k}]) is thus
$\epsilon_{\kappa}\,=\,{\frac{3\,\epsilon}{4}}$. The ellipticity of
the potential $\epsilon_{\psi}$, obtained similarly by taking moments
of $\psi$, is $\epsilon_{\psi} = \epsilon/4$.  Note that the
ellipticity of the potential $\epsilon_{\phi}$ computed above is
smaller than $\epsilon$ (by a factor of 4), as it is weighted by the
circular gaussian window function. Comparing, the weighted
ellipticities, we find that $\epsilon_{\kappa} > \epsilon_{\psi}$, as
expected since equi-potentials are always rounder than the mass
contours.

\section{Measuring the flattening of the mass distribution}
\label{flattening}
We now show how these results can be used to measure the flattening of
the mass distribution. As in ordinary galaxy-galaxy lensing, the
galaxy catalog is separated into a foreground and a background
subsample, using magnitude, colors or photometric redshifts.  The
ellipticity of the galaxies in both subsamples is then measured in the
by taking second moments of the light distribution. The ellipticities
of the foreground sample yields the ellipticity of the light
$\epsilon_{l}$ associated with each lens. While $\epsilon_{l}$ is
ignored in ordinary galaxy-galaxy lensing, we instead align the
foreground galaxies along their major axes before stacking. We then
measure the average ellipticity of the mass $\epsilon_{\kappa}$ as
described above, by replacing the integrals in
equations~(\ref{eq:q_shear}) and (\ref{eq:t_m_shear}) by sums over the
sheared background galaxies. This yields a measurement of the
component of the average ellipticity of the mass, $\epsilon_{\kappa
\parallel}$ parallel to the that of the light, i.e.
\begin{equation}
\epsilon_{\kappa \parallel} = {\rm Re} \langle \epsilon_{\kappa}^{*} 
\widehat{\epsilon}_{l} \rangle,
\end{equation}
where the ellipticities are taken to be complex numbers with $\epsilon=
\epsilon_{1}+i\epsilon_{2}$, $^{*}$ denotes complex conjugation,
and $\widehat{\epsilon}_{l} = \epsilon_{l}/|\epsilon_{l}|$ is the
unit ellipticity of the light.

We now compute the uncertainty in measuring $\epsilon_{\kappa
\parallel}$, by taking the square of the mean of the discrete
estimators for $M$, $T$ and $Q_{\parallel}={\rm Re}(Q)$, and
converting back into integrals (Schneider \& Bartelmann 1997). 
In the absence of lensing, we find
\begin{eqnarray}
\sigma^2[M] & = & \frac{\sigma_{\epsilon}^{2}}{n_{b}
n_{f} A} \int d^{2}x~ h_{r}^{2}(x), \nonumber \\
\sigma^2[T] & = & \frac{\sigma_{\epsilon}^{2}}{n_{b}
n_{f} A} \int d^{2}x~ g_{r}^{2}(x), \nonumber \\
\sigma^2[Q_{\parallel}] & = & \frac{\sigma_{\epsilon}^{2}}{2 n_{b}
n_{f} A} \int d^{2}x~ \left[ g_{r}^{2}(x) + g_{\times}^{2}(x) \right],
\end{eqnarray}
where $\sigma_{\epsilon}^{2}=\langle \epsilon_{r}^{2} \rangle =
\langle \epsilon_{\times}^{2} \rangle$ is the variance of the
intrinsic ellipticity distribution of galaxies ($\sim 0.3^{2}$),
$n_{b}$ and $n_{f}$ are respectively the number density of background
and foreground galaxies, and $A$ is the area covered by the survey.

For the elliptical isothermal model with the gaussian weight function,
we can evaluate these integrals and find,
\begin{eqnarray}
\sigma^{2}[M] & = & \frac{\sigma_{\epsilon}^{2}}{4 \pi n_{b} n_{f} A
\beta^{2}},~~~
\sigma^{2}[T] =  \frac{\sigma_{\epsilon}^{2} \beta^{2}}{2\pi n_{b}
n_{f} A}, \nonumber \\
\sigma^{2}[Q_{\parallel}] & = & \frac{3 \sigma_{\epsilon}^{2}
\beta^{2}}{4 \pi n_{b} n_{f} A}
\end{eqnarray}
By propagating these errors in the definition of the ellipticity of
the mass 
(Eq.~[\ref{eq:epsilon_k}]), we can compute the signal to
noise ratio $\left( S/N \right)_{\epsilon_{\kappa}} = \epsilon_{\kappa
\parallel}/\sigma[\epsilon_{\kappa \parallel}]$ for measuring
$\epsilon_{\kappa \parallel}$, and find, to first order in 
$\epsilon$,
\begin{eqnarray}
\left(\frac{S}{N}\right)_{\epsilon_{\kappa}}& \simeq
&4.6\,\left(\frac{\epsilon_{\kappa}}{0.3} \right)
\left(\frac{\alpha}{0.5''}\right) \left(\frac{n_{b}}{1.5 \,{\rm
arcmin^{-2}}} \right)^{\frac{1}{2}} \\ \nonumber &\,& \left(
\frac{n_f}{0.035 {\rm arcmin^{-2}}} \right)^{\frac{1}{2}}\,
\left(\frac{0.3}{\sigma_{\epsilon}} \right)\, \left(\frac{\rm A}{1000
{\rm deg}^{2}} \right)^{\frac{1}{2}}.
\end{eqnarray}
Here, we have chosen to scale $\epsilon_{\kappa}$ in units of 0.3,
which is reasonable given current observational limits on the
flattening of dark matter halos (see Table 3 in the Sackett (1999)
review and references therein; Buote \& Canizares 1998).
Additionally, in these scalings, we have used the survey
specifications (ellipticity dispersion, number density of foreground
lenses $n_f$ and the number density of background galaxies $n_b$, and
approximate observed Einstein radius) quoted for the SDSS
commissioning run provided by Fischer et al. (1999), with a modestly
expanded area (1000 deg$^{2}$) from the current area of 225 deg$^{2}$.
In the context of estimating the scatter arising in the mass estimates
from galaxy-galaxy lensing due to halo shapes, Fischer et al. (2000)
mention in passing that with 10 times more data than the commissioning
run, halo shapes can be measured; this figure is comparable to our
estimate of the $S/N$.

Note that since these numbers have been taken from the SDSS
commissioning run (which suffered from poor seeing), are conservative
and do take into account several observational errors. The dispersion
of 0.3 in the ellipticity distribution includes a correction for the
seeing (see for instance weak lensing observations of Rhodes et
al. (2000) and Bacon et al. (2000) and references therein). The effect
of seeing and noise are also reflected in the modest number density
assumed for background sources. Note, however, that in contrast to the
case of ordinary galaxy-galaxy lensing, foreground galaxies will need
to be sufficiently resolved to measure their shapes prior to alignment
and stacking. This could induce some degradation in the signal, but
since the foreground galaxies are typically brighter and larger, this
effect is expected to be small.

The shape parameters of the mass will therefore be easily detectable
with SDSS in the near future (McKay et al. 2000). For the total SDSS
area of $10^{4}$ deg$^{2}$, the significance rises to 15$\sigma$. This
in fact implies that potentially, even the radial dependence of the
flattening can be studied by considering annuli-shaped weight
functions (for instance, the difference of two gaussians). Moreover,
the degree of flattening as a function of the morphological galaxy
type can also be studied.

It is interesting to compare the $(S/N)_{\epsilon_{\kappa}}$ expected
for measuring $\epsilon_{\kappa \parallel}$ estimated above with that
of the usual galaxy-galaxy lensing $(S/N)_{M} = M / \sigma[M]$ which
measures the mass enclosed within an aperture. For the model
considered here, we find the following relation,
\begin{eqnarray}
\left( \frac{S}{N}
\right)_{\epsilon_{\kappa}}\,=\,0.17\,
\left( \frac{\epsilon_{\kappa}}{0.3} \right)
\, \left( \frac{S}{N} \right)_{M}.
\end{eqnarray}
Therefore, shape parameters can be measured with a significance which
is smaller but comparable to that of the enclosed mass. Note that for
the current SDSS survey area of $A=225$ deg$^{2}$, we find $(S/N)_{M}
\simeq 13$ which agrees roughly with the significance of the reported
Fischer et al. (1999) results, when averaged over all radial
bins. 

\section{Measuring the alignment of light and mass}
\label{alignment}

Given the good prospects expected from the above results, one can be
more ambitious and try to characterize the alignment of mass and light
in more detail. We can make use of the amplitude of the light
ellipticity $\epsilon_{l}$, which we have not used until now. This can
be done by grouping the lens galaxies into several
$\epsilon_{l}$-bins, and computing $\epsilon_{\kappa \parallel}$
separately for each bin. A more direct approach would be to consider
the correlation function of the ellipticities of the mass and light,
defined as
\begin{equation}
C_{\kappa l} = {\rm Re} \langle \epsilon_{\kappa}^{*} \epsilon_{l} \rangle,
\end{equation}
where the average is over all lens galaxies, and $\epsilon_{\kappa}$
is an estimate of the mass ellipticity of each lens derived from its
associated background galaxies. While $\epsilon_{\kappa}$ for an
individual lens is rather noisy, a significant measurement of
$C_{\kappa l}$ can be obtained by averaging over a large number of
lens galaxies. This correlation function could also be computed for
several annuli, and, would therefore, yield a direct measure of the
radial dependence of the alignment of mass and light.

\section{Discussion}
\label{discussion}

The shapes of dark matter halos (see Sackett (1999) for a more
comprehensive review) have been probed via many techniques and the
consensus from these studies is that the precise shapes offer
important clues to both the galaxy formation process and perhaps, even
to the nature of dark matter. Cosmological N-body simulations suggest
that dark matter halos are triaxial and that dissipation 
determines their shape. High resolution simulations
find that the effect of dissipation (Katz \& Gunn 1991; Dubinski 1994)
is to transform an initially triaxial halo from prolate-triaxial to
oblate-triaxial, while preserving the degree of flattening, yielding
on average rounder and more oblate dark halos than those in
dissipationless simulations.

Comparing the shape of the mass profile inferred from X-ray data for a
sample of ellipticals with that of the light, Buote \&
Canizares (1994; 1998) find that the dark matter is at least as
flattened as the light and is definitely more extended. The origin of
the X-ray isophotal twist in the case of NGC720, they argue reflects an
intrinsic misalignment of the stars with the dark matter.  Keeton,
Kochnek \& Seljak (1997) incorporate the shape of the light 
distribution as a constraint in modeling individual lenses (that
produce multiple images of background quasars) and find that an
additional component of shear is required to match the
observations. They speculate that this component could arise from a
misalignment between the luminous galaxy and its dark matter halo.
Our proposed technique will provide reliable measurements of 
the shape and orientation of light and mass in galaxies, thereby
aiding in the understanding of the coupling of baryons with the dark 
matter.

\acknowledgements

We thank Martin Rees and Donald Lynden-Bell for useful discussions.
PN acknowledges support from a Trinity College Research Fellowship 
and AR from the EEC Lensing Network for a TMR post-doctoral 
fellowship and a Wolfson College Research Fellowship.

\end{document}